\documentclass[twocolumn,showpacs,jcp,superscriptaddress]{revtex4}
\usepackage{graphicx}
\usepackage[usenames]{color}
\usepackage{amssymb}
\usepackage{amsmath}
\usepackage{amsfonts}
\usepackage{mathrsfs}

\renewcommand{\epsilon}{\varepsilon}
\newcommand{\figurewidth}{0.46\textwidth}

\begin{document}
\title{Ejection dynamics of a ring polymer out of a nanochannel}

\author{Junfang Sheng}
\affiliation{CAS Key Laboratory of Soft Matter Chemistry, Department of Polymer Science and Engineering, University of Science and Technology of China, Hefei, Anhui Province 230026, P. R. China}

\author{Kaifu Luo}
\altaffiliation[]{Author to whom the correspondence should be addressed}
\email{kluo@ustc.edu.cn}
\affiliation{CAS Key Laboratory of Soft Matter Chemistry, Department of Polymer Science and Engineering, University of Science and Technology of China, Hefei, Anhui Province 230026, P. R. China}

\date{\today}

\begin{abstract}

We investigate the ejection dynamics of a ring polymer out of a
cylindrical nanochannel using both theoretical analysis and three
dimensional Langevin dynamics simulations. The ejection dynamics for
ring polymers shows two regimes like for linear polymers, depending
on the relative length of the chain compared with the channel. For
long chains with length $N$ larger than the critical chain length
$N_{c}$, at which the chain just fully occupies the nanochannel, the
ejection for ring polymers is faster compared with linear chains of
identical length due to a larger entropic pulling force; while for
short chains ($N<N_c$), it takes longer time for ring polymers to
eject out of the channel due to a longer distance to be diffused to
reach the exit of the channel before experiencing the entropic
pulling force. These results can help understand many biological
processes, such as bacterial chromosome segregation.

\end{abstract}


\maketitle

\section{Introduction}

The properties of a polymer confined in a nanochannel have attracted
broad interest \cite{Brochard,Kremer,Sotta,Cifra,Gong,Klushin,Yang}
because they are of fundamental relevance in polymer physics and are
also related to many biological processes, such as double-stranded
DNA genomes packaging inside the phage capsid \cite{Smith}, polymers
transport through nanopore \cite{Luo1,Luo2} and viruses injecting
their DNA into a host cell \cite{Miller}.

The importance of cyclic structures in biological macromolecular
science is strikingly demonstrated by the existence of circular DNA,
cyclic peptides and cyclic oligosaccharides and polysaccharides
\cite{Cyclic}. Ring closure of a polymer is one of the important
factors influencing its statistical mechanical properties.
Understanding the static and dynamic properties of ring polymer is a
challenging problem due to the difficulties inherent to a systematic
theoretical analysis of such objects constrained to a unique
topology. The scaling behavior of isolated, highly diluted, ring
polymers has been studied. des Cloizeaux \cite{Cloizeaux}, Deutsch
\cite{Deutsch} and Grosberg \cite{Grosberg} discussed the effect of
topological constraints on the properties of ring polymers, and
found that the topological constraint and the excluded volume have
similar effects. The radius of gyration for large single ring polymers obey
the same scaling relationship as that of linear chains \cite{Deutsch,Grosberg}, although
this is not true for ring polymers in a melt or ring polymer brushes \cite{Halverson,Sakaue,Reith}.

Ring closure acts as an important role in a wide range of biophysical contexts where DNA is constrained: segregation of the compacted circular genome of some bacteria \cite{Jun}, formation of chromosomal territories in cell nuclei \cite{Dorier}, compaction and ejection of the knotted DNA of a virus \cite{Marenduzzo,Matthews}, migration of a circular DNA in an electrophoresis gel \cite{Obukhov} or in a nanochannel \cite{Reisner}.

After three decades of intensive research, the conformational
properties of a self-avoiding polymer chain confined in a slit or in
a cylindrical nanochannel are relatively well understood.
\cite{Daoud,Gennes,Milchev10,Milchev11,Arnold}. However, a deeper
understanding of the basic properties of ring polymer in confined
environments is a field in its infancy \cite{Persson,Witz}.
Only few studies have addressed semiflexible ring polymers. Ostermeir \textit{et al.} \cite{Frey} investigated the internal structure of semiflexible ring polymers in weak spherical confinement and found buckling and a conformational transition to a figure eight form. Fritsche and Heermann \cite{Heermann} examined the conformational properties of a semiflexible ring polymer confined to different geometrical constraints and found that the geometry of confinement plays a important role in shaping the spatial organization of polymers. Most recently, we have found the helix chain conformation of flexible ring polymers confined to a cylindrical nanochannel, and demonstrated that the longitudinal size along the channel for a ring polymer scales as $N\sigma(\sigma/D)^{2/3}$, the same as that for a linear chain but with different prefactors. Here $D$ is the radius of the channel, $N$ the chain length and $\sigma$ the Kuhn length of the chain \cite{Sheng}. We further gives the theoretical ratio value 0.561 of the longitudinal size for a ring polymer and a linear chain of the same $N$.

As to the dynamics of the polymer under confinements, Milchev \textit{et al.} \cite{Milchev10} have investigated the ejection of linear chain out of nanopore using Monte carlo simulation and found that the ejection dynamics depends on the chain length.
Unlike its linear polymer counterpart, the dynamics of confined ring polymers is still lacking, although many bimolecules are circular.
To this end, in this work we study the ejection dynamics of a ring polymer confined in a nanochannel by means of analytical techniques and Langevin dynamics simulations.
The basic questions associated with this process are the following:
(a) what's effect of the chain length and the channel length on the
ejection dynamics? (b) what's the difference of the ejection
dynamics for ring polymers compared with the linear one? For a fixed channel, which one is faster compared a ring polymer with a linear chain of the identical length?

We believe that this work is interesting and important for understanding biological systems with more complexity, such as viruses injecting their DNA into a host cell, the behavior of DNA inside phages or the spatial organization of the bacterial nucleoid in \textit{E. coli}.

\section{Model and methods} \label{chap-model}

In our numerical simulations, the polymer chains are modeled as bead-spring chains of Lennard-Jones (LJ) particles with the Finite Extension Nonlinear
Elastic (FENE) potential. Excluded volume interaction between beads is modeled by a short range repulsive LJ potential: $U_{LJ}
(r)=4\epsilon [{(\frac{\sigma}{r})}^{12}-{(\frac{\sigma} {r})}^6]+\epsilon$ for $r\le 2^{1/6}\sigma$ and 0 for $r>2^{1/6}\sigma$. Here, $\sigma$ is the diameter of
a bead, and $\epsilon$ is the depth of the potential. The connectivity between neighboring beads is modeled as a FENE spring with $U_{FENE}(r)=-\frac{1}{2}kR_0^2\ln(1-r^2/R_0^2)$, where $r$ is the distance between consecutive beads, $k$ is the spring constant and $R_0$ is the maximum allowed separation between connected beads.

\begin{figure}
\includegraphics*[width=\figurewidth]{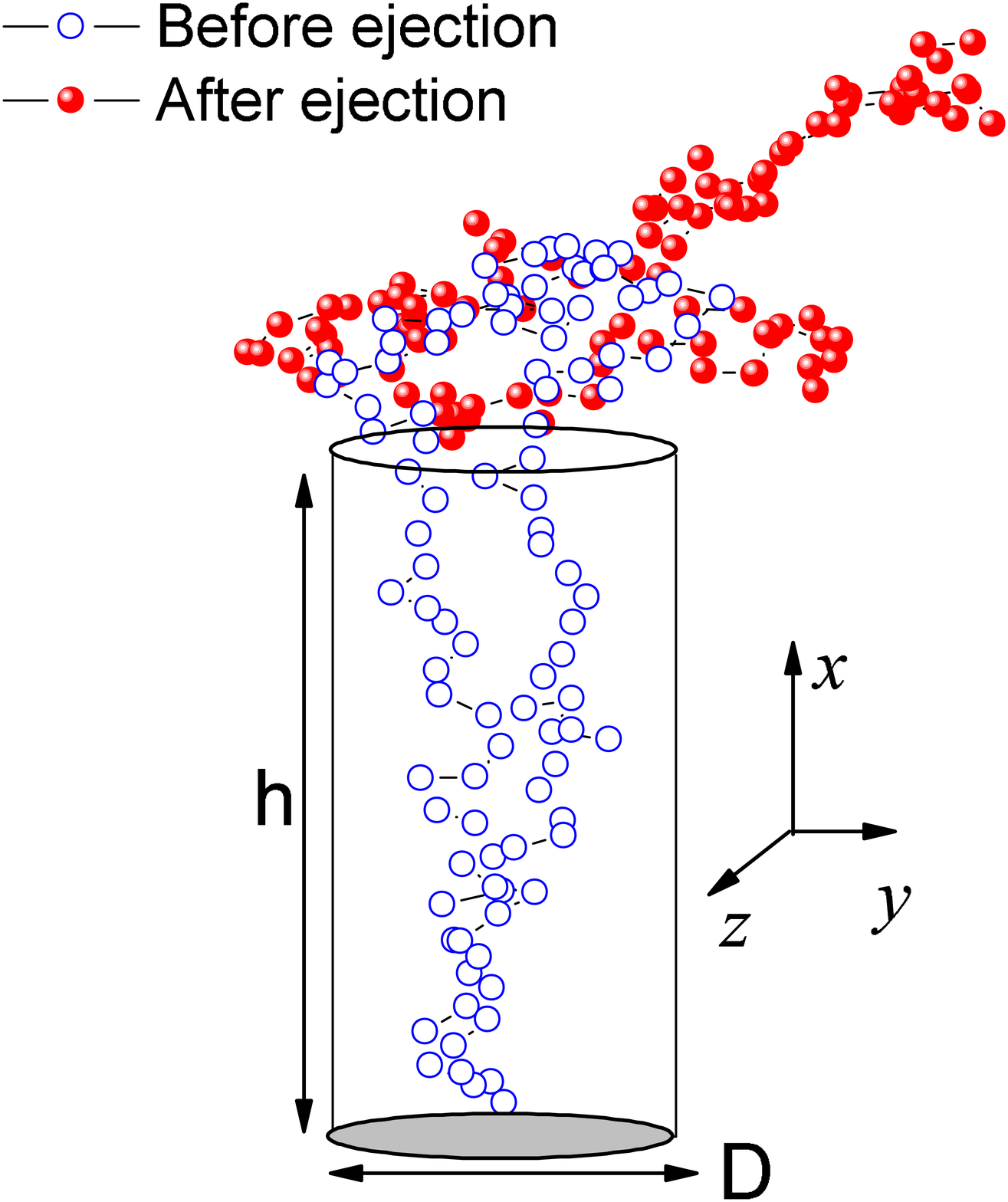}
\caption{A schematic representation of the initial
and the finial conformation of a ring polymer ejection from a
cylindrical channel in three dimensions. Here, the channel diameter
$D=5$, the channel height $h=20.5$ and the polymer length $N=100$.
        }
 \label{Fig1}
\end{figure}

We consider a schematic representation as shown in Fig. \ref{Fig1},
where a ring polymer is confined in a cylindrical channel with one
end sealed. The nanochannel and the sealed surface are described by
stationary particles within distance $\sigma$ from one another which
interact with the beads by the repulsive Lennard-Jones potential.
The particle positions of the nanochannel and the sealed surface are
not changed in the simulations.

In the Langevin dynamics simulation, each bead is subjected to
conservative, frictional, and random forces, respectively, with
\cite{Allen} $m{\bf \ddot {r}}_i =-{\bf
\nabla}({U}_{LJ}+{U}_{FENE})-\xi {\bf v}_i+{\bf F}_i^R$. Here $m$ is
the bead's mass, $\xi$ is the friction coefficient, ${\bf v}_i$ is
the bead's velocity, and ${\bf F}_i^R$ is the random force which
satisfies the fluctuation-dissipation theorem. In the present work,
the LJ parameters $\epsilon$, $\sigma$, and $m$ fix the system
energy, length and mass units respectively, leading to the
corresponding time scale $t_{LJ}=(m\sigma^2/\epsilon)^{1/2}$ and
force scale $\epsilon/\sigma$, which are of the order of ps and pN,
respectively. The dimensionless parameters in the model are then
chosen to be $R_0=1.5$, $k=15$, $\xi=0.7$.

In our model, each bead corresponds to a Kuhn length (twice of the persistence length) of a polymer. For a single-stranded DNA (ssDNA), the persistence length of the ssDNA is sequence and solvent dependent and varies in a wide range, to our knowledge, usually from about 1 to 4 nm. We assume the value of $\sigma \sim 2.8$ nm for a ssDNA containing approximately four nucleotide bases. The average mass of a base in DNA is about 312 amu, so the bead mass $m \approx 1248$ amu. We set $k_{B}T=1.2\epsilon$, which means that the interaction strength $\epsilon$ is $3.39 \times 10^{-21}$ J at actual temperature 295 K. This leads to a time scale of 69.2 ps and a force scale of 1.2 pN. The Langevin equation is then integrated in time by a method described by Ermak and Buckholz \cite{Ermak}.

We initially fix the last monomer of the linear chain but anyone of the ring polymer at the sealed bottom of the nanochannel, while the remaining monomers are under thermal collisions described by the Langevin thermostat to obtain an equilibrium configuration. In order to learn the mechanism of chain ejection out of the nanochannel, the link of the monomer with the bottom of the channel is removed, then the chain is released to diffuse along the channel. The residence time $\tau$ is measured, once all monomers pass the opening at $x=h$ and leave the channel.  Typically, we average our data over 700 independent runs.

\section{Results and discussion} \label{chap-results}
\subsection{Scaling arguments}

\subsubsection{Longitudinal size of a polymer in infinitely long nanochannel}

According to the blob picture, for a linear polymer confined in a
infinitely long three-dimensional nanochannel of diameter $D$, the
chain will extend along the channel axis forming a string of blobs
of size $D$. The center of the blob is on the axis of the
nanochannel. For each blob, $D=Ag^\nu\sigma $ due to the dominant
excluded volume effects, where $g$ is the number of monomers in a
blob, $\sigma$ is the Kuhn length of the chain, $\nu$ is the Flory
exponent in three dimensions, and $A$ is a constant. Thus, each blob
contains $g=(\frac{D}{A\sigma})^{\frac{1}{\nu}}$ monomers, and the
number of blobs is $n_b=N/g=N
(\frac{A\sigma}{D})^{\frac{1}{\nu}}$.
The free energy cost for the chain confinement is proportional to
the number of blobs, thus the free energy in units of $k_BT$ is
$\mathcal{F}=B_lN(A\sigma/D)^{1/\nu}$, with $B_l$ being a constant.
The blob picture then predicts the longitudinal size of the linear chain to be $R_{{\parallel},l}=n_bD = ND(\frac{A\sigma}{D})^{\frac{1}{\nu}}=(A\sigma)^{
\frac{1}{\nu}}ND^{1-\frac{1}{\nu}}$.
Using $\nu=3/5$ in three dimensions, we obtain
\begin{equation}
R_{\parallel,l}=(A\sigma)^{\frac{5}{3}}ND^{-\frac{2}{3}}.
\label{eq1}
\end{equation}

In order to model the the chain conformation for a ring polymer confined in a nanochannel, we have extended the blob picture \cite{Sheng}. For a ring polymer, the chain will extend along the channel axis forming two strings of blobs of $D/2$,the two strings of blobs show helix structure. For each blob of size $D/2$, $D/2=Ag_r^\nu\sigma$ with $g_r$ being the the number of monomers in a blob. Here, the same prefactor $A$ for ring polymers and linear chains is due to the same solution environment.
Thus, each blob contains $g_r=(\frac{D}{2A\sigma})^{\frac{1}{\nu}}$ monomers, and the number of blobs is $n_b=N/g_r=N(\frac{2A\sigma}{D})^{\frac{1}{\nu}}$. The free energy cost in units of $k_BT$ is $\mathcal{F}=B_rN(\frac{2A\sigma}{D})^{\frac{1}{\nu}}$, with $B_r$ being a constant.

By geometrical analysis, the distance between two successive layers is $\frac{\sqrt{2}}{4}D$, and so the total length occupied by blobs in the channel is  $R_{\parallel,r}=\frac{\sqrt{2}}{4}D(\frac{N}{2g_r}-1)+\frac{1}{4}D+\frac{1}{4}D=
D(\frac{\sqrt{2}}{8}\frac{N}{g_r}+\frac{1}{2}-\frac{\sqrt{2}}{4})$. When $\frac{N}{g_r}$ is very large, $R_{\parallel,r} \approx \frac{\sqrt{2}}{8}D\frac{N}{g_r}=
\frac{\sqrt{2}}{8}DN(\frac{A\sigma}{D/2})^{\frac{1}{\nu}}=\frac{\sqrt{2}}{8}2^{\frac{1}{\nu}}(A\sigma)^{\frac{1}{\nu}} ND^{1-\frac{1}{\nu}}$.
Using $\nu=3/5$ in three dimensions, we obtain
\begin{equation}
R_{\parallel,r}=\frac{\sqrt{2}}{8}2^{\frac{5}{3}}(A\sigma)^{\frac{5}{3}} ND^{-\frac{2}{3}}.
\label{eq2}
\end{equation}

Therefore, the longitudinal size along the channel for a ring
polymer scales as $R_{\parallel}\sim N\sigma(\sigma/D)^{2/3}$, the
same as that for a linear chain but with a different prefactor.
The ratio of the longitudinal sizes along the nanochannel (or the
prefactors) for a ring polymer and a linear chain is
\begin{equation}
\frac{R_{\parallel,r}}{R_{\parallel,l}}=\frac{\sqrt{2}}{8}2^{\frac{5}{3}}=0.561.
\label{eq3}
\end{equation}
If using more accurate value of $\nu=0.588$, we have $\frac{R_{\parallel,r}}{R_{\parallel,l}}=0.575$.

The simulation results \cite{Sheng} confirm the above predictions and give $(A\sigma)^{1/\nu}=1.367\pm0.009$ for the parameters used in the model.

\subsubsection{Ejection dynamics of a polymer confined in a nanochannel}

Intuitively, for the ejection of a polymer out of a nanochannel, the dynamics is controlled by the relative length of polymer compared with the channel height $h$. There exists a critical polymer length $N_c$, where the polymer just fully occupies the channel, namely $R_{\parallel}(N_c)=h$. Thus, for linear chains the critical length $N_{c,l}$ is
\begin{equation}
N_{c,l}=\frac{h}{(A\sigma)^{\frac{5}{3}}D^{-\frac{2}{3}}},
\label{eq4}
\end{equation}
while for ring polymers the critical length $N_{c,r}$ reads
\begin{equation}
N_{c,r}=\frac{4\sqrt{2}h}{2^{\frac{5}{3}}(A\sigma)^{\frac{5}{3}} D^{-\frac{2}{3}}}=\frac{1.782h}{(A\sigma)^{\frac{5}{3}} D^{-\frac{2}{3}}}.
\label{eq5}
\end{equation}
The ratio of the critical length for the ring polymers and the linear polymer is
\begin{equation}
\frac{N_{c,r}}{N_{c,l}}=\frac{4\sqrt{2}}{2^{\frac{5}{3}}}=1.782.
\label{eq6}
\end{equation}

Short chains with chain length of $N<N_c$ are initially fully confined in the nanochannel while long polymers with chain length of $N>N_c$ initially occupy
the whole channel with several segments outside the channel exit.
For long chains with $N>N_c$, the ejection is a driven process where the pulling force $f$ is from the entropy and is induced by already ejected monomers \cite{Klushin}. For short chains with $N<N_c$, polymer needs to move to the channel exit by a diffusive process, and then experiences a pulling force
as for long chains.

We assume the ejection process to be quasi-equilibrium. For long chains, the pulling force can be estimated from the free energy $\mathcal{F}$ of a chain partially
confined in the nanochannel with the innermost monomer being at distance $x$ from the channel exit.
For long linear chains, $x=n(t)(A\sigma)^{1/\nu}D^{1-1/\nu}$ and the free energy $\mathcal{F}=B_l n(t)(A\sigma/D)^{1/\nu}k_BT=B_l\frac{k_BT}{D}x$, with $n(t)$ being
the number of monomers inside the channel at time $t$.
The differential of the free energy allows an estimate of the pulling force
\begin{equation}
f_l=B_l\frac{k_BT}{D}.
\label{eq7}
\end{equation}
It is worthy of noting that $f$ is independent of the tail length as well as $h$, but inversely proportional to
$D$.
For long ring chains, $x=n(t)\frac{\sqrt{2}}{8}2^{\frac{1}{\nu}}(A\sigma)^{\frac{1}{\nu}}D^{1-\frac{1}{\nu}}$ and the free energy $\mathcal{F}=B_r n(t)(2A\sigma/D)
^{1/\nu}k_BT=4\sqrt{2}B_r\frac{k_BT}{D}x$.
Thus, the pulling force
\begin{equation}
f_r=4\sqrt{2}B_r\frac{k_BT}{D}.
\label{eq8}
\end{equation}

We further have the ratio of the pulling force for long ring polymers and linear chains
\begin{equation}
\frac{f_r}{f_l}=\frac{4\sqrt{2}B_r}{B_l},
\label{eq9}
\end{equation}
which is only determined by the universal prefactors for ring polymers and linear chains.

During the ejection process, the pulling force induced by the tail is balanced by the total friction. Namely, for long chains we have
\begin{equation}
\xi n(t)\frac{dx}{dt}=-f,
\label{eq10}
\end{equation}
where $\xi$ is the friction coefficient per monomer. Taking into account the relationship
of $x(t)$ and $n(t)$, we obtain the ejection time
\begin{equation}
\tau_{long,l}= \frac{\xi h^2}{2f_l(A\sigma)^{1/\nu}D^{1-1/\nu}}=\frac{\xi h^2D^{1/\nu}}{2B_l(A\sigma)^{1/\nu}k_BT}
\label{eq11}
\end{equation}
for long linear chains, and
\begin{equation}
\tau_{long,r}= \frac{2\sqrt{2}\xi h^2}{f_r(2A\sigma)^{1/\nu}D^{1-1/\nu}}=\frac{\xi h^2D^{1/\nu}}{2B_r(2A\sigma)^{1/\nu}k_BT}
\label{eq12}
\end{equation}
for long ring polymers.
Therefore, the ratio of the ejection time for long ring polymers and linear chains is
\begin{equation}
\frac{\tau_{long,r}}{\tau_{long,l}}=\frac{B_l}{B_r2^{1/\nu}}=0.315\frac{B_l}{B_r},
\label{eq13}
\end{equation}
where $\nu=3/5$ is used.

As noted above, for short polymers ($N<N_{c}$), it undergoes a diffusive process before the first segment exiting the channel, and subsequently the ejection process driven by a pulling force.
Accordingly, we divide the total ejection time $\tau$ into two parts, $\tau_{1}$ for the diffusive process and $\tau_{2}$ for the driven process.

For the the diffusive process of short linear chains, $\tau_1$ is
\begin{equation}
\tau_{1,l}=\frac{(h-R_{\parallel,l})^2}{2D_{diff}}=\frac{N\xi[h-(A\sigma)^{\frac{1}{\nu}}ND^{1-\frac{1}{\nu}}]^2}{2k_BT},
\label{eq14}
\end{equation}
with $D_{diff}=\frac{k_BT}{N\xi}$ being the diffusion constant. In
addition, for the driven process $\tau_2$ can be written as
\begin{equation}
\tau_{2,l}=\frac{\xi R_{\parallel,l}^2D^{1/\nu}}{2B_l(A\sigma)^{1/\nu}k_BT}=\frac{\xi N^2(A\sigma)^{\frac{1}{\nu}}D^{2-\frac{1}{\nu}}}{2B_l k_BT}.
\label{eq15}
\end{equation}
Here, $\tau_{2,l}$ is negligible compared to $\tau_{1,l}$ for quite short chains, and then the ejection time $\tau_l\approx\tau_{1,l}$.
Based on the differential of the ejection time with $N$, $\frac{\partial\tau_l}{\partial
N}=[h-(A\sigma)^{\frac{1}{\nu}}ND^{1-\frac{1}{\nu}}][h-3(A\sigma)^{\frac{1}{\nu}}ND^{1-\frac{1}{\nu}}]$=0, we obtain  the critical chain length $N_{c,l}=\frac{h}
{(A\sigma)^{\frac{1}{\nu}}D^{1-\frac{1}{\nu}}}$ as shown in Eq. (\ref{eq4}) and another resolution
\begin{equation}
N^{\ast}_l=N_{c,l}/3=\frac{h}{3(A\sigma)^{\frac{1}{\nu}}D^{1-\frac{1}{\nu}}},
\label{eq16}
\end{equation}
where the residence time $\tau_l$ reaches to its maximum value
\begin{equation}
\tau_{max,l}=\frac{2\xi h^3}{27(A\sigma)^{\frac{1}{\nu}}D^{1-\frac{1}{\nu}}k_BT}.
\label{eq17}
\end{equation}

For the the diffusive process of short ring chains, $\tau_1$ is
\begin{equation}
\tau_{1,r}=\frac{(h-R_{\parallel,r})^2}{2D_{diff}}=\frac{N\xi[h-\frac{\sqrt{2}}{8}2^{\frac{1}{\nu}} (A\sigma)^{\frac{1}{\nu}}ND^{1-\frac{1}{\nu}}]^2}{2k_BT},
\label{eq18}
\end{equation}
In addition, for the driven process $\tau_2$ can be written as
\begin{equation}
\tau_{2,r}=\frac{\xi R_{\parallel,r}^2D^{\frac{1}{\nu}}}{2B_r(2A\sigma)^{1/\nu}k_BT}=\frac{\xi N^2(2A\sigma)^{\frac{1}{\nu}}D^{2-\frac{1}{\nu}}}{64B_rk_BT}.
\label{eq19}
\end{equation}
Again, $\tau_{2,r}$ is negligible compared to $\tau_{1,r}$ for quite
short chains, and then $\tau_r\approx\tau_{1,r}$. Based on the
differential of the ejection time with $N$,
$\frac{\partial\tau_r}{\partial N}=[h-\frac{\sqrt{2}}{8}(2A\sigma)^{\frac{1}{\nu}}ND^{1-\frac{1}{\nu}}][h-\frac{3\sqrt{2}}{8}(2A\sigma)^{\frac{1}{\nu}}ND^{1-\frac{1}{\nu}}]=0$,
we obtain  the critical chain length
$N_{c,r}=\frac{4\sqrt{2}h}{(2A\sigma)^{\frac{1}{\nu}}D^{1-\frac{1}{\nu}}}$
as in Eq. (\ref{eq5}) and another resolution
\begin{equation}
N^{\ast}_r=N_{c,r}/3=\frac{4\sqrt{2}h}{3(2A\sigma)^{\frac{1}{\nu}}D^{1-\frac{1}{\nu}}},
\label{eq20}
\end{equation}
where the ejection time $\tau$ reaches to its maximum value
\begin{equation}
\tau_{max,r}=\frac{8\sqrt{2}\xi h^3}{27(2A\sigma)^{\frac{1}{\nu}}D^{1-\frac{1}{\nu}}k_BT}.
\label{eq21}
\end{equation}

Thus, we have
\begin{equation}
\frac{N^{\ast}_r}{N^{\ast}_l}=\frac{\tau_{max,r}}{\tau_{max,l}}=\frac{4\sqrt{2}}{2^{\frac{1}{\nu}}}=1.782.
\label{eq22}
\end{equation}
%
%
%

\subsection{Simulation results}

\begin{figure}
\includegraphics*[width=\figurewidth]{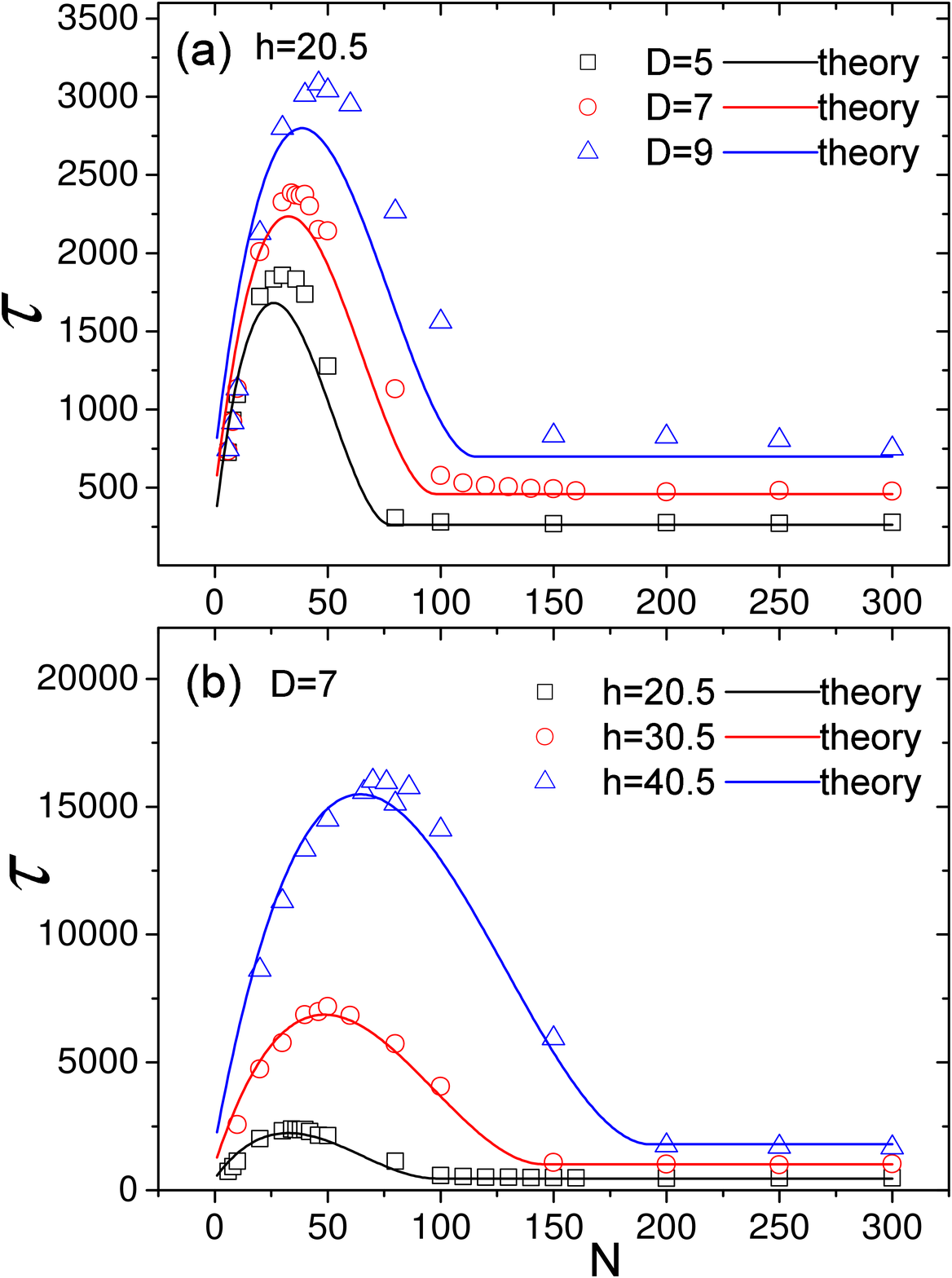}
\caption{(a) Average ejection time $\tau$ as a
function of the polymer length $N$ for the channel height $h=20.5$ and
different channel diameters. (b) The same for channel diameter $D=7$ and different channel heights. The full lines are plotted using Eqs. (\ref{eq18}) and (\ref{eq19}).
        }
\label{Fig2}
\end{figure}

\begin{figure}
\includegraphics*[width=\figurewidth]{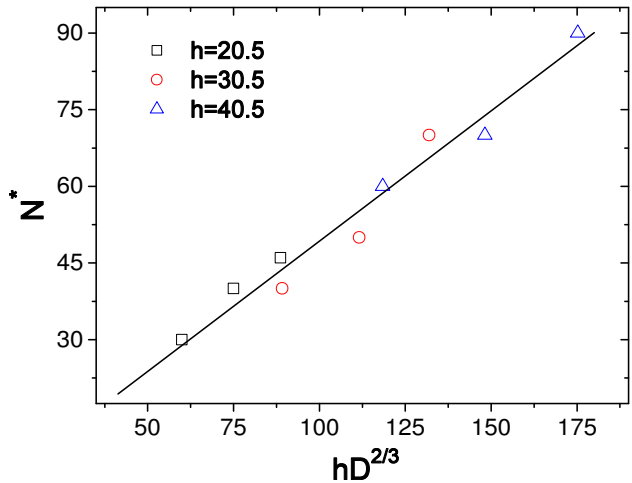}
\caption{Scaling of $N_{c}$ at which the ejection
time owns its maximum as a function of $hD^{2/3}$ for different $h$ and $D$. Here, $h=20.5$, 30.5, and 40.5 and $D=5$, 7 and 9.
        }
\label{Fig3}
\end{figure}

\begin{figure}
\includegraphics*[width=\figurewidth]{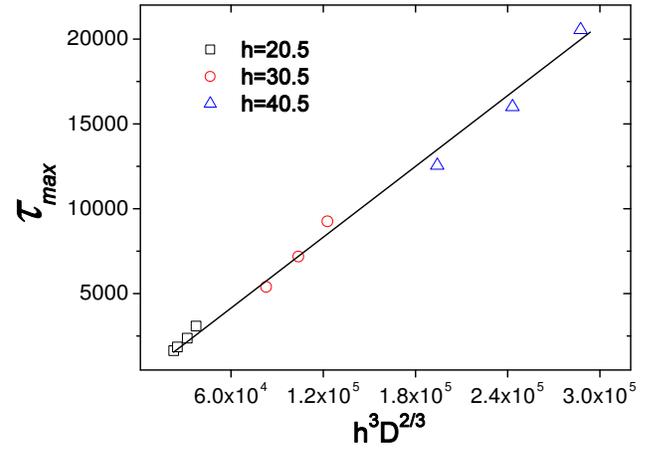}
\caption{Scaling of the maximum of ejection time of
different polymer sizes as a function of $h^3D^{2/3}$ for different
$h$ and $D$. Here, we use $D=5$, 7 and 9 for $h=30.5$ and 40.5, and $D=4$, 5,
7 and 9 for $h=20.5$.
        }
\label{Fig4}
\end{figure}

The average ejection time $\tau$ as a function of the ring polymer length $N$ for different channel diameters ($D=5, 7$ and 9) at fixed channel height $h=20.5$ and for different channel heights
($h=20.5, 30.5$, and 40.5) at channel diameter $D=7$ are shown in  Fig. \ref{Fig2}a and  Fig. \ref{Fig2}b, respectively. The two pictures show that ejection time increases with the increase of channel diameter and channel height.
Moreover, we get a special polymer length $N^{\ast}$ at which the ejection time meets its maximum.  Fig. \ref{Fig3} shows the plot of $N^{\ast}_r$ against $hD^{2/3}$ for different $D$ and $h$. All the data points collapse on the same line, which is in agreement with the prediction in Eq. (\ref{eq20}). The line plotted in  Fig. \ref{Fig4} proves the prediction in Eq. (\ref{eq21}).


\begin{figure}
\includegraphics*[width=\figurewidth]{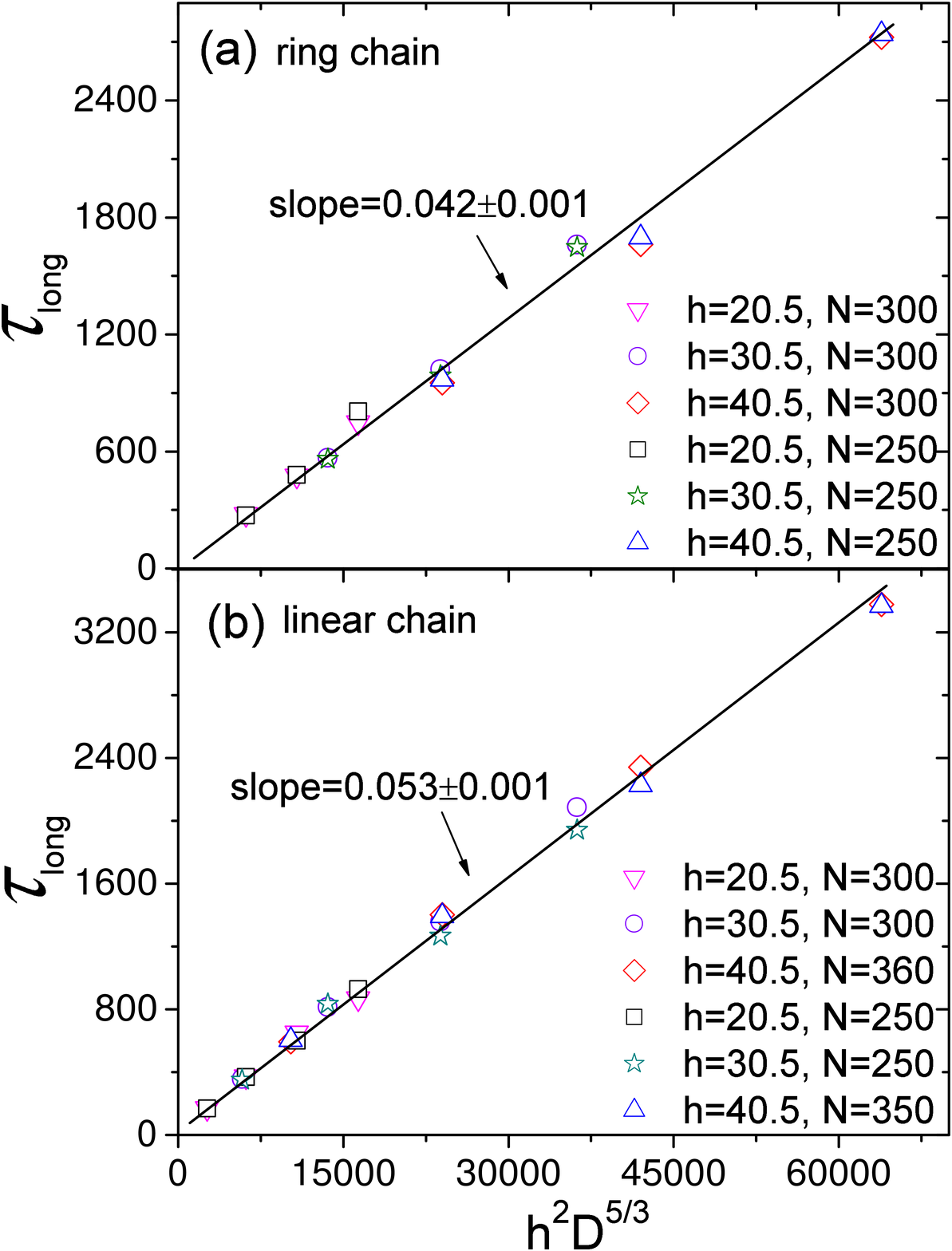}
\caption{ The ejection time $\tau_{long}$ as a function of
$h^2D^{5/3}$: (a) ring chain with $N=250, 300$ ($D=5$, 7 and 9 for
$h=20.5$, 30.5 and 40.5); (b) linear chain with $N=250$, 300 ($D=3$, 5 and 7 for $h=20.5$ and 30.5), $N=350$, and 360 ($D=9$ for $h=40.5$).
        }
\label{Fig5}
\end{figure}

As noted before, there exists a critical polymer length $N_c$ at which the polymer just fully occupies the channel. Short chains ($N<N_c$) are initially fully confined in the nanochannel while long polymers ($N>N_c$) initially occupy the whole channel with several segments outside the channel exit. From the platforms in Fig. \ref{Fig2}, we obtain the ejection time $\tau_{long}$ for long polymers ($N>N_c$).

Fig. \ref{Fig5}a and Fig. \ref{Fig5}b show the scaling plot of
$\tau_{long}$ with $h^2D^{5/3}$ for both ring polymers and linear
chains, respectively. For different polymer lengths, channel heights
and channel diameters, all data points collapse on the same line in
Fig. \ref{Fig5}a and Fig. \ref{Fig5}b, respectively. These
results confirm the predictions in Eqs. (\ref{eq11}) and
(\ref{eq12}). In addition, the slopes are 0.042 and 0.053 for ring
polymer and linear chain,  respectively. This indicates
$\frac{\tau_{long,r}}{\tau_{long,l}}=\frac{0.042}{0.053}=0.792$.
Based on Eqs. (\ref{eq11}) and (\ref{eq12}), we have $B_r=1.60$,
$B_l=4.02$ and thus $\frac{B_r}{B_l}=0.398$ using the parameters
$\xi=0.7$, $T=1.2$ and $(A\sigma)^{1/\nu}=1.367$. Moreover, we
further obtain $\frac{f_r}{f_l}=2.250$ through Eq. (\ref{eq9}), which
demonstrates that the driving force induced by confinement for long
ring polymers is larger than that for linear chains.
Using Eqs. (\ref{eq18}) and (\ref{eq19}) to fit curves in Fig. \ref{Fig2}, we find that the numerical results are qualitatively described by theoretical findings.

\begin{figure}
\includegraphics*[width=\figurewidth]{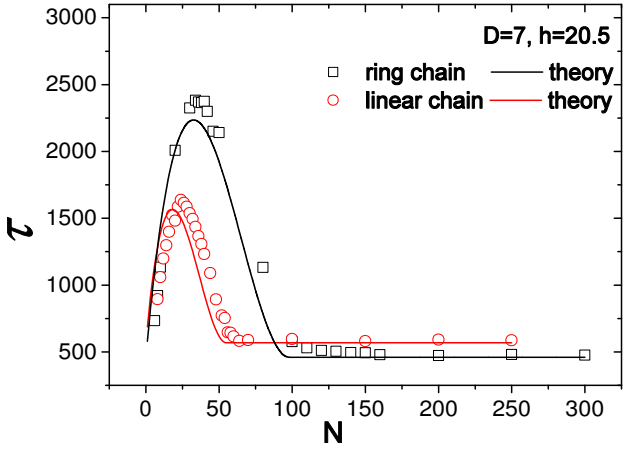}
\caption{Average ejection time as a function of the polymer length for the
linear chain and the ring polymer. Here, $D=7$, and $h=20.5$. The full line plotted to fit the data points for linear chain is based on Eqs. (\ref{eq14}) and (\ref{eq15}).
        }
\label{Fig6}
\end{figure}

\begin{figure}
\includegraphics*[width=\figurewidth]{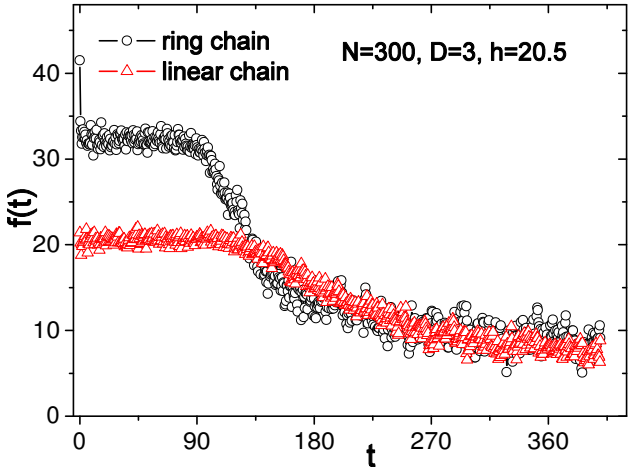}
\caption{The force $f$ exerted on the innermost monomer of the ring
polymer and the linear chain. Here, $N=300$, $D=3$, and $h=20.5$.
        }
\label{Fig7}
\end{figure}

\begin{figure}
\includegraphics*[width=\figurewidth]{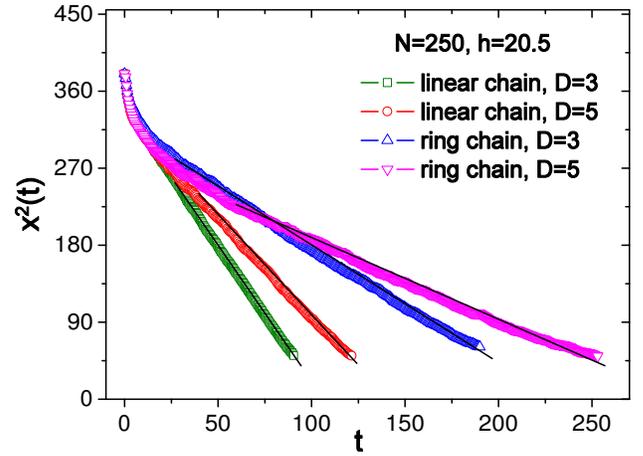}
\caption{Squared distance $x^2(t)$ of the last monomer from the
channel exit during the ejection as a function of the time for the ring polymer and the linear
chain and different channel diameters. Here, $h=20.5$ and $N=250$.
        }
\label{Fig8}
\end{figure}

To compare the ejection dynamics for ring polymers with that for linear chains, we show the ejection time as a function of the chain length $N$ for $D=7$ and $h=20.5$ in Fig. \ref{Fig6}.
One does see characteristic differences: for short chains ($N<N_c$), it takes longer time for ring polymers to eject out of the channel than that for linear chains; while for long chains ($N>N_c$), linear chains need longer time. These findings are in agreement with the predictions in Eqs. (\ref{eq13}) and (\ref{eq22}). Ring polymers has smaller $R_{\parallel}$ than that for linear chains of the same $N$ and thus ring polymers must diffuse longer distance to reach the exit of the channel. When the chain length is larger than the critical chain length ($N>N_c$), the force exerted on the residual segments for ring polymer is larger than that for linear chain due to the smaller blob size in the channel for ring polymers than that for linear chains as predicted $\frac{f_r}{f_l}=2.250$.
The platform of the force at small time $t$ shown in Fig. \ref{Fig7} for both ring polymer and linear chain confirms this prediction.

In Fig. \ref{Fig6}, we find $\frac{N^{\ast}_r}{N^{\ast}_l}=\frac{34}{24}=1.417$,
$\frac{\tau_{max,r}}{\tau_{max,l}}=\frac{2383.031}{1637.134}=1.456$,
which are predicted as 1.782. The difference may be from the
non-equilibrium process of the ejection. In addition,
$\frac{N_{c,r}}{N_{c,l}}=\frac{110}{64}=1.719$ and it is
predicted as 1.782 in Eq. (\ref{eq6}).

The mean-squared distance $x^2(t)$ of the last monomer with respect to the channel exit against elapsed time after the release of the last monomer is shown in Fig. \ref{Fig8} for both ring polymers and linear chains. The lines plotted according to the curves are based on the equation $x^2(t)=x^2(0)-\frac{2f_l(A\sigma)^{1/\nu}D^{1-1/\nu}} {\xi} t$ for a linear polymer and the equation $x^2(t)=x^2(0)-\frac{f_r(2A\sigma)^{1/\nu}D^{1-1/\nu}} {2\sqrt{2}\xi} t$ for a ring polymer, which indicates that the Eq. (\ref{eq10}) can correctly describe the ejection dynamics.

\begin{figure}
\includegraphics*[width=\figurewidth]{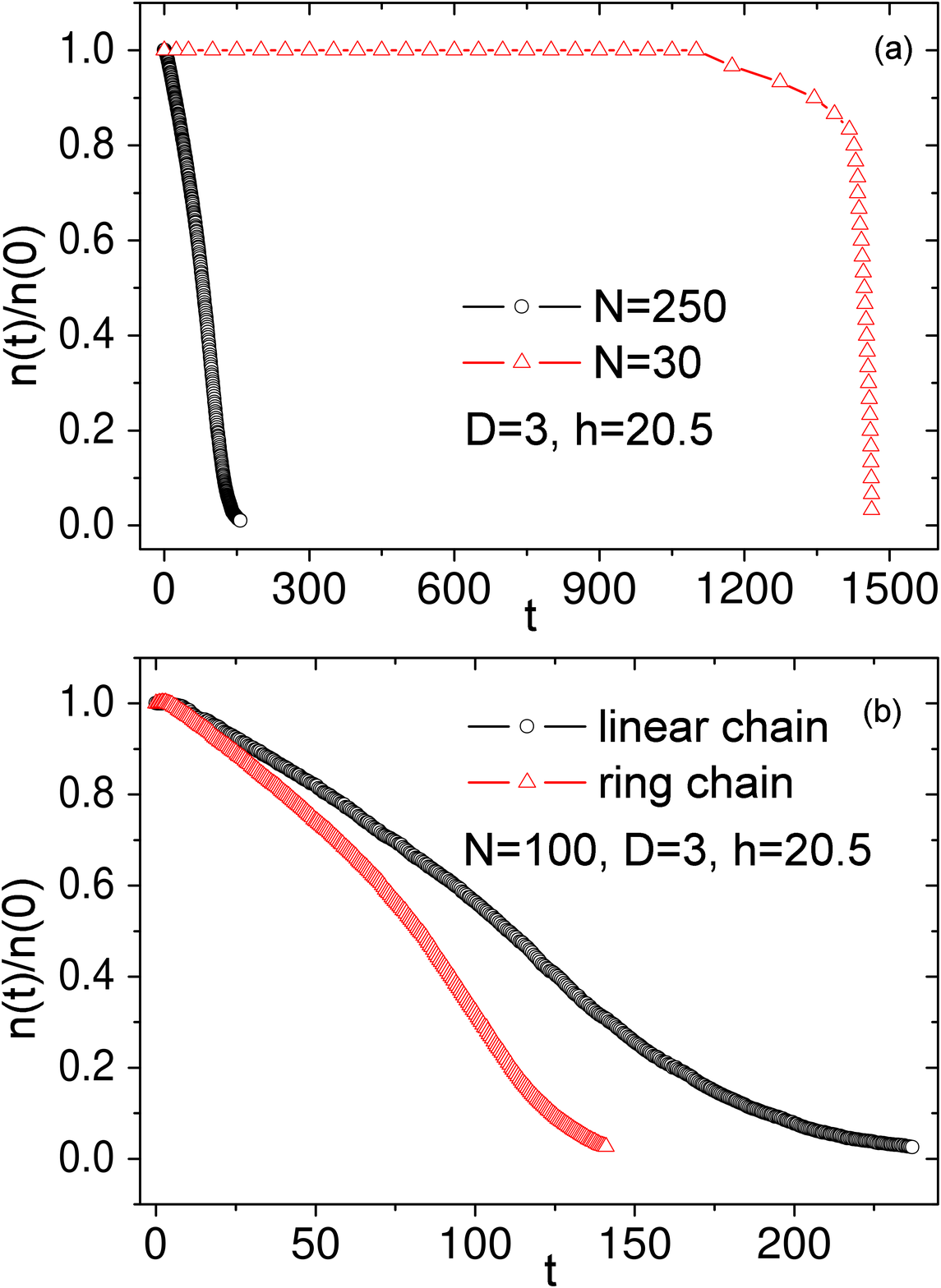}
\caption{Numbers of monomers $n(t)$ inside the channel normalized by its value at $t=0$ as a function of the time for (a) ring polymers with different lengths; (b) the ring polymer and the linear chain of length $N=100$. Here, $D=3$ and $h=20.5$.
        }
\label{Fig9}
\end{figure}

\begin{figure}
\includegraphics*[width=\figurewidth]{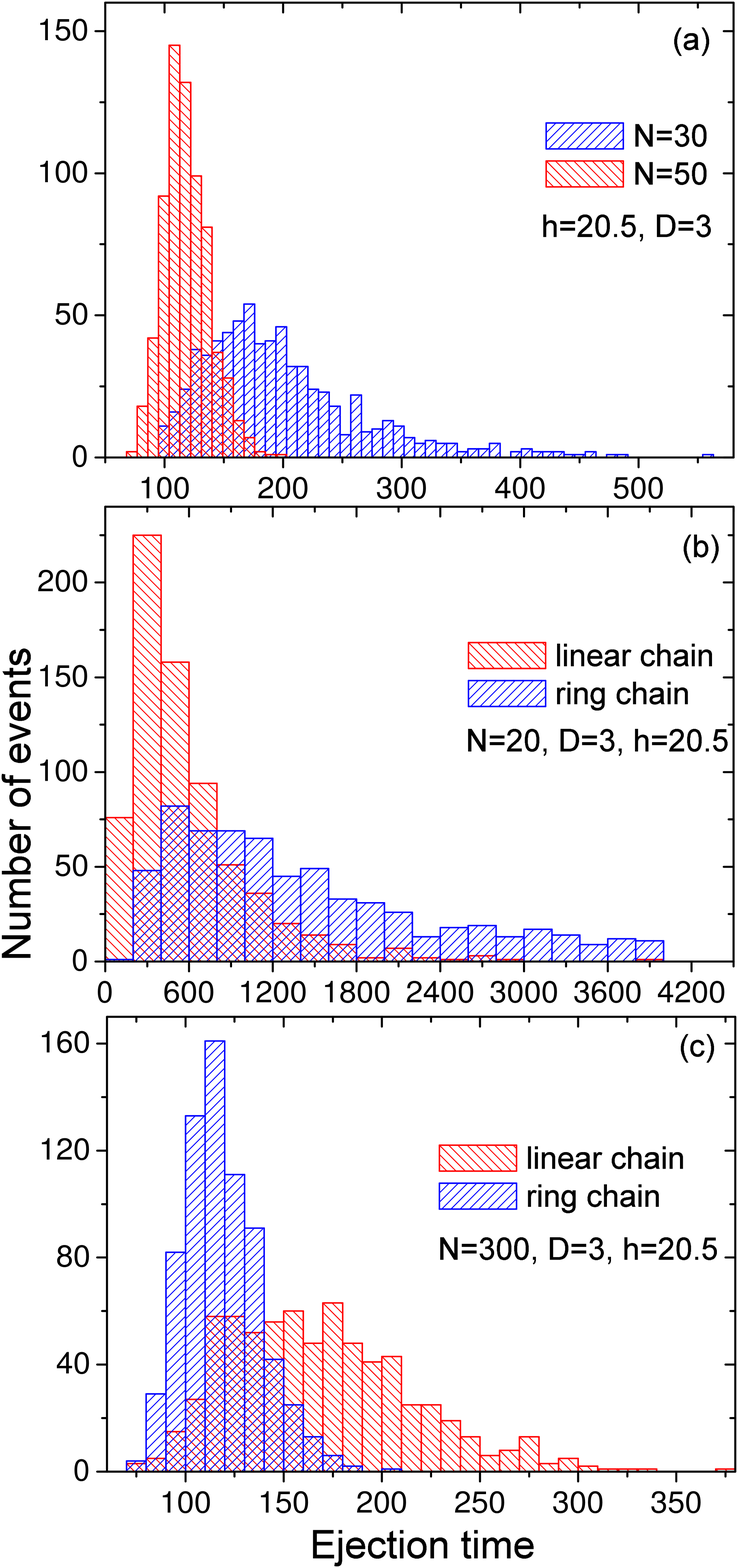}
\caption{The distribution of the ejection time: (a) ring polymers of different chain lengths; (b) the ring polymer and the linear chain of length $N=20$; (c) the ring polymer
and the linear chain of length $N=300$. Here, $D=3$ and $h=20.5$.
        }
\label{Fig10}
\end{figure}

In order to know the details in the ejection process, we record the number of residual monomers inside the channel in the total process, $n(t)/n(0)$ (normalized by its value at $t=0$). We see that the short ring polymer experiences a diffusion process before it starts to eject out of the channel, corresponding to the platform in the plot for $N=30$ as shown in Fig. \ref{Fig9}a. When the chain length $N>N_c$, the ejection process is faster for the ring polymer than that for a linear one, which can also be inferred from the portion of residual monomers at time $t$, as presented in  Fig. \ref{Fig9}b.

Fig. \ref{Fig10}a shows the histograms of the ejection time for
ring polymers with different chain lengths. The ejection time
distribution for polymer of length $N=30$ has a long tail and is
much wider than that for $N=50$.
The ejection time distributions for ring polymers and linear chains at both short and long chain lengths are given in Fig. \ref{Fig10}b and Fig. \ref{Fig10}c, respectively.
For short chain $N=20$, it takes longer time for ring polymer to leave the channel than that for linear chain, and the ejection time distribution for the ring polymer is wider and has a long tail. For long chains $N=300$, however, the result is opposite, reflecting the larger driving force for the ring polymer than that for the linear chain.

Nature not only imposes geometrical constraints on biopolymers by
confinement through cell membrane, the cell nucleus or viral capsid,
but also exploits the advantages of certain underlying chain
topologies, such as the ring structure. In fact, \textit{E. coli}
has a rod-shaped geometry and its chromosome is not a linear polymer
but a circular one. Based on Monte Carlo simulations, Jun and Mulder
\cite{Jun} addressed a basic physical issue associated with
bacterial chromosome segregation in rod-shaped cell-like geometry.
By simulations of two ring polymers, in the same setting as the
linear ones and they found that two ring polymers segregate more
readily than linear ones in confinement. According to our above
theoretical analysis and simulation results, for ring polymers
confined in a cylindrical nanochannel the blob size for ring
polymers is smaller than that for linear polymers, which indicates
that during the chromosome segregation the driving force for ring
polymers is larger than that for linear one, leading to faster
segregation.


\section{Conclusions} \label{chap-conclusions}

We investigate the ejection dynamics of a ring polymer out of a cylindrical nanochannel using both theoretical analysis and three dimensional Langevin dynamics simulations. The ejection dynamics for ring polymers shows two regimes like for linear polymers, depending on the relative length of the chain compared with the channel. For long chains with length $N$ larger than the critical chain length $N_c$, at which the chain just fully occupies the nanochannel, the ejection for ring polymers is faster compared with linear chains of identical length due to a larger entropic pulling force; while for short ($N<N_c$), it takes longer time for ring polymers to eject out of the channel due to a longer distance to be diffused to reach the exit of the channel before experiencing the entropic pulling force. These results can help understand many biological processes.

Our results should enable a new understanding of the conformational statistics
and dynamics of confined ring biopolymers such as DNA. The concrete graph about ring polymer confined in a nanochannel needs more deep study so as to realize many complex problems in both
biochemistry and theoretical study. Our findings are of interest for
(molecular) biological/biochemical, technology as well as physics reasons.

\begin{acknowledgments}
This work is supported by the National Natural Science Foundation of China (Grant No. 21074126, 21174140), the Specialized Research Fund for the Doctoral Program of Higher
Education (Grant No. 20103402110032), and the ``Hundred Talents Program'' of Chinese Academy of Science (CAS).
\end{acknowledgments}

\includegraphics*[width=8cm]{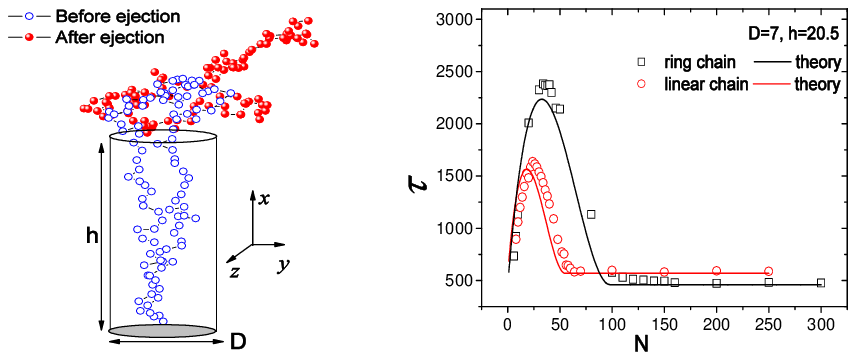}

\begin{thebibliography}{8}
\bibitem{Brochard} F. Brochard-Wyart, P. G. de Gennes, \textit{J. Chem. Phys.} {\bf 67}, 52 (1977).
\bibitem{Kremer} K. Kremer, K. Binder, \textit{J. Chem. Phys.} {\bf 81}, 6381 (1984).
\bibitem{Sotta} P. Sotta, A. Lesne, J. M. Victor, \textit{J. Chem. Phys.} {\bf 112}, 1565 (2000).
\bibitem{Cifra} P. Cifra, \textit{J. Chem. Phys.} {\bf 131}, 224903 (2009).
\bibitem{Gong} Y. Gong, Y. Wang, \textit{Macromolecules.} {\bf 35}, 7492 (2002).
\bibitem{Klushin} L. I. Klushin, A. M. Skvortsov, H. P. Hsu, K. Binder, \textit{Macromolecules.} {\bf 41}, 5890 (2008).
\bibitem{Yang} Y. Yang, T. M. Burkhard, G. Gompper, \textit{Phys. Rev. E.} {\bf 76}, 011804 (2007).
%
\bibitem{Smith} D. E. Smith, S. J. Tans, S. B. Smith, S. Grimes, D. L. Anderson, and C. Bustamante, \textit{Nature} {\bf 413}, 748 (2001).
\bibitem{Luo1} K. Luo, T. Ala-Nissila, S. C. Ying,  \textit{J. Chem. Phys.} {\bf 124}, 034714 (2006).
\bibitem{Luo2} K. Luo, T. Ala-Nissila, S. C. Ying, \textit{Phys. Rev. Lett.} {\bf 100}, 058101 (2008).
\bibitem{Miller} R. V. Miller, \textit{Sci. Am.} {\bf 278}, 66 (1998).
\bibitem{Cyclic} J. A. Semlyen, \textit{Cyclic Polymers}, 2nd ed. (Springer, Dordrecht, 2000).
\bibitem{Cloizeaux} J. des Cloizeaux, \textit{J. Phys. Lett.} {\bf 42}, 433 (1981).
\bibitem{Deutsch} J. M. Deutsch, \textit{Phys. Rev. E.} {\bf 59}, R2539 (1999).
\bibitem{Grosberg} A. Y. Grosberg, \textit{Phys. Rev. Lett.} {\bf 85}, 3858 (2000).
%
\bibitem{Halverson} J. D. Halverson, W. B. Lee, G. S. Grest,  A. Y. Grosberg, and K. Kremer, \textit{J. Chem. Phys.} {\bf 134}, 204904 (2011).
\bibitem{Sakaue} T. Sakaue, \textit{Phys. Rev. Lett.} {\bf 106}, 167802 (2011).
\bibitem{Reith} D. Reith, A. Milchev, P. Virnau, and K. Binder, \textit{EPL} {\bf 95}, 28003 (2011).
%
\bibitem{Jun} S. Jun and B. Mulder, \textit{Proc. Natl. Acad. Sci. U.S.A.} {\bf 103}, 12388 (2006).
\bibitem{Dorier} J. Dorier and A. Stasiak, \textit{Nucleic Acids Res.} {\bf 37}, 6316 (2009).
\bibitem{Marenduzzo} D. Marenduzzo and C. Micheletti, \textit{J. Mol. Biol.} {\bf 330}, 485 (2003).
\bibitem{Matthews} R. Matthews, A. A. Louis, and J. M. Yeomans, \textit{Phys. Rev. Lett.} {\bf 102}, 088101 (2009).
\bibitem{Obukhov} S. P. Obukhov, M. Rubinstein, and T. Duke, \textit{Phys. Rev. Lett.} {\bf 73}, 1263 (1994).
\bibitem{Reisner} W. Reisner, K. J. Morton, R. Riehn, Y. M. Wang, Z. Yu, M. Rosen, J. C. Sturm, S. Y. Chou, E. Frey, and R. H. Austin1, \textit{Phys. Rev. Lett.} {\bf 94}, 196101 (2005).
\bibitem{Daoud} M. Daoud and P. G. de Gennes, \textit{J. Physique} {\bf 38}, 85 (1977).
\bibitem{Gennes} P. G. de Gennes, \textit{Scaling Concepts in Polymer Physics} (Cornell University Press, Ithaca, NY, 1979).
\bibitem{Milchev10} A. Milchev, L. Klushin, A. Skvortsov and K. Binder, \textit{Macromolecules} {\bf 43}, 6877 (2010).
\bibitem{Milchev11} A. Milchev, \textit{J. Phys.: condens. Matter} {\bf 23}, 103101 (2011).
\bibitem{Arnold} A. Arnold, B. Bozogui, D. Frenkel, B. Y. Ha, and S. Jun, \textit{J. Chem. Phys.} {\bf 127}, 164903 (2007).
%
\bibitem{Persson} F. Persson, P. Utko, W. Reisner, N. B. Larsen, and A. Kristensen, \textit{Nano Lett.} {\bf 9}, 1382 (2009).
\bibitem{Witz} G. Witz, K. Rechendorff, J. Adamcik, and G. Dietler, \textit{Phys. Rev. Lett.} {\bf 106}, 248301 (2011).
\bibitem{Frey} K. Ostermeir, K. Alim, and E. Frey, \textit{Phys. Rev. E} {\bf 81}, 061802 (2010); \textit{Soft Matter} {\bf 6}, 3467 (2010)
\bibitem{Heermann} M. Fritsche, and D. Heermann, \textit{Soft Matter} {\bf 7}, xxx (2011), See Doi: 10.1039/c1sm05445g.
\bibitem{Sheng} J. Sheng and K. Luo, to be published.
%
\bibitem{Allen} M. P. Allen and D. J. Tildesley, \textit{Computer Simulation of Liquids} (Oxford University, New York, 1987)
\bibitem{Ermak} D. L. Ermak and H. Buckholz, \textit{J. Comput. Phys.} {\bf 35}, 169 (1980).


\end{thebibliography}
\end{document}